\documentclass[12pt]{article}
\usepackage{amsmath,amssymb,dsfont}
\usepackage{graphicx}

\newtheorem{theorem}{Theorem}

\newtheorem{corollary}[theorem]{Corollary}

\newtheorem{example}[theorem]{Example}

\newtheorem{lemma}[theorem]{Lemma}

\newtheorem{remark}[theorem]{Remark}

\def\Cbb{\mathbb{C}}

\def\diag{{\rm diag}}

\def\half{\frac{1}{2}}

\def\Rbb{\mathbb{R}}
\def\Cbb{\mathbb{C}}

\def\diag{{\rm diag}}

\begin{document}

\title{Synthesis of linear quantum stochastic systems via quantum feedback networks\footnote{Research supported by the
Australian Research Council (ARC).}}

\author{Hendra I. Nurdin
\thanks{H.~I.~Nurdin is with the Department of Engineering, College of Engineering
and Computer Science, The Australian National University,
Canberra, ACT 0200, Australia. Email: Hendra.Nurdin@anu.edu.au.}
}
\date{}
\maketitle

\begin{abstract}
Recent theoretical and experimental investigations of coherent feedback control, the feedback control of
a quantum system with another quantum system, has raised the
important problem of how to synthesize a class of quantum systems,
called the class of linear quantum stochastic systems, from basic
quantum optical components and devices in a systematic way. The
synthesis theory sought in this case can be naturally viewed as a
quantum analogue of linear electrical network synthesis theory and
as such has potential for applications beyond the realization of
coherent feedback controllers. In earlier work, Nurdin, James and
Doherty have established that an arbitrary linear quantum
stochastic system can be realized as a cascade connection of
simpler one degree of freedom quantum harmonic oscillators, together
with a direct interaction Hamiltonian which is bilinear in the
canonical operators of the oscillators. However, from an
experimental perspective and based on current methods and
technologies, direct interaction Hamiltonians are challenging to
implement for systems with more than just a few degrees of
freedom. In order to facilitate more tractable physical realizations
of these systems, this paper develops a new synthesis algorithm
for linear quantum stochastic systems that relies solely on
field-mediated interactions, including in implementation of the
direct interaction Hamiltonian.  Explicit synthesis examples are
provided to illustrate the realization of two degrees of freedom
linear quantum stochastic systems using the new algorithm.
\end{abstract}

\section{Introduction}
\label{sec:intro} Linear quantum stochastic systems (see, e.g.,
\cite{EB05,JNP06,NJP07b,NJD08}) arise as an important class of
models in quantum optics \cite{GZ00} and in phenomenological
models of quantum RLC circuits \cite{YD84}. They are used, for
instance, to model optical cavities driven by coherent laser
sources and are of interest for applications in quantum
information science. In particular, they have the potential as a
platform for realization of entanglement networks
\cite{Man06,MW07,YNJP08} and to function as sub-sytems of a
continuous variable quantum information system (e.g., the scheme
of \cite{LB99}). Recently, there has also been interest in the
possibilities of control of linear quantum stochastic systems with
a controller which is a quantum system of the same type
\cite{YK03b,JNP06,NJP07b,Mab08} (thus involving no measurements of
quantum signals), often referred to as  ``coherent-feedback control'', and an experimental realization of
a coherent-feedback control system for broadband disturbance attenuation has
been successfully demonstrated by Mabuchi \cite{Mab08}. Linear
quantum stochastic systems are a particularly attractive class of
quantum systems to study for coherent control because of their
simple structure and complete parametrization by a number of
matrix parameters.

A natural and important question that arose out of the studies on
coherent control is how an arbitrary linear quantum stochastic
system can be built or synthesized in a systematic way, in the
quantum optical domain, from a bin of quantum optical components
like beam splitters, phase shifters, optical cavities, squeezers,
etc; this can be viewed as being analogous to the question in
electrical network synthesis theory of how to synthesize linear
analog circuits from basic electrical components like resistors,
capacitors, inductors, op-amps, etc. The synthesis problem is not
only of interest for quantum control, but is a timely subject in
its own right given the current intense research efforts in
quantum information science (see, e.g., \cite{DM03}). These
developments present significant opportunities for investigations
of a network synthesis theory (e.g., \cite{AV73} for linear
electrical networks) in the quantum domain as a significant
direction for future development of circuit and systems theory. In
particular, such a quantum synthesis theory may be especially
relevant for the theoretical foundations, development and  design
of future linear photonic integrated circuits.
%In particular, these circuits have the potential to play a role as a key sub-system of a continuous
%variable, possibly all-optical, quantum information processing system.

As a first step in addressing the quantum synthesis question,
Nurdin, James and Doherty \cite{NJD08} have shown that any linear
quantum stochastic system can, in principle, be synthesized by a
cascade of simpler one degree of freedom quantum harmonic
oscillators together with a direct interaction Hamiltonian between
the canonical operators of these oscillators. Then they also
showed how these one degree of freedom harmonic oscillators and
direct bilinear interaction Hamiltonian can be synthesized from
various quantum optical components. However, from an experimental
point of view, direct bilinear interaction Hamiltonians between
independent harmonic oscillators are challenging to implement
experimentally with current technology for systems that have more
than just a few degrees of freedom (possibly requiring some
complex spatial arrangement and orientation of the oscillators)
and therefore it becomes important to investigate approximate
methods for implementing this kind of interaction. Here we propose
such a method by exploiting the recent theory of quantum feedback
networks \cite{GJ08}. In our scheme, the direct interaction
Hamiltonians are approximately realized by appropriate field
interconnections among oscillators. The approximation is based on
the assumption that the time delays required in establishing field
interconnections are vanishingly small, which is typically the
case in quantum optical systems where quantum fields propagate at
the speed of light.

The paper is organized as follows. Section
\ref{sec:linear-summary}  provides a brief overview of linear
quantum stochastic systems and the concatenation and series
product of such sytems. Section \ref{sec:model-matrix} recalls the
notion of a model matrix and the concatenation of model matrices,
while Section \ref{sec:edges} recalls the notion of edges, their
elimination, and reduced Markov models. Section \ref{sec:prior} reviews
a prior synthesis result from \cite{NJD08}. Section
\ref{sec:main-results} presents the main results of this paper and
an example to illustrate their application to the realization of a
two degrees of freedom linear quantum stochastic system. In Section
\ref{sec:passive}  the special class of passive linear quantum stochastic systems
is introduced and it is shown that such systems can be synthesized by using
only passive sub-systems and components. Another synthesis example for a passive
system is also presented. Finally,
Section \ref{sec:conclu} offers the conclusions of this paper. In order to
focus on the results, all proofs are collected together in the
Appendix.

\section{Linear quantum stochastic systems}
\label{sec:linear-summary} This section serves to recall some
notions and results on linear quantum stochastic systems that are
pertinent for the present paper. A relatively detailed overview of
linear quantum stochastic systems can be found in \cite{NJD08} and
further discussions in \cite{EB05,JNP06,Nurd07,NJP07b}, thus they
will not be repeated here.

Throughout the paper we shall use the following notations:
$i=\sqrt{-1}$, $^*$ will denote the adjoint of a linear operator
as well as the conjugate of a complex number, if $A=[a_{jk}]$ is a
matrix of linear operators or complex numbers then
$A^{\#}=[a_{jk}^*]$, and $A^{\dag}$ is defined as
$A^{\dag}=(A^{\#})^T$, where $^T$ denotes matrix transposition. We
also define, $\Re\{A\}=(A+A^{\#})/2$ and $\Im\{A\}=(A-A^{\#})/2i$
and denote the identity matrix by $I$ whenever its size can be
inferred from context and use $I_{n}$ to denote an $n \times n$
identity matrix. Similarly, $0$ denotes  a matrix with zero
entries whose dimensions can be determined from context, while
$0_{m \times n}$ denotes a matrix with specified dimension $m
\times n$ with zero entries. Other useful notations that we shall employ is
$\diag(M_1,M_2,\ldots,M_n)$ (with $M_1,M_2,\ldots,M_n$ square
matrices) to denote a block diagonal matrix with
$M_1,M_2,\ldots,M_n$ on its diagonal block, and $\diag_{n}(M)$
($M$ a square matrix) denotes a block diagonal matrix with the
matrix $M$ appearing on its diagonal blocks $n$ times.

Let $q_1,p_1,q_2,p_2,\ldots,q_n,p_n$ be the canonical position and
momentum operators of a {\em many degrees of freedom quantum
harmonic oscillator} satisfying the canonical commutation
relations (CCR) $[q_j,p_k]=2i\delta_{jk}$. The integer $n$ will be
referred to as the {\em degrees of freedom} of the oscillator.
Letting $x=(q_1,p_1,q_2,p_2,\ldots,q_n,p_n)^T$ then these
commutation relations can be written compactly as:
$$
xx^T-(xx^T)^T=2i\Theta,
$$
with $\Theta=\diag_{n/2}(J)$ and $J=\left[\begin{array}{cc} 0 & 1
\\-1 & 0 \end{array} \right]$. Here a {\em linear quantum stochastic
system} $G$ is a quantum system defined by three {\em parameters}:
(i) A quadratic Hamiltonian $H=\half x^T R x$ with $R=R^T \in
\Rbb^{n \times n}$, (ii) a coupling operator $L=Kx$ where $K$ is
an $m \times n$ complex matrix and (iii) a unitary $m \times m$
scattering matrix $S$. We also assume that the system oscillator
is in an initial state with density operator $\rho$. For
shorthand, we write $G=(S,L,H)$ or $G=(S,Kx,\half x^TRx)$. The
time evolution, in the interaction picture,  $X(t)$ of $x$ ($ t
\geq 0$) is given by the quantum stochastic differential equation
(QSDE):
\begin{align}
dX(t)&=AX(t)dt+B\left[\begin{array}{c} dA(t)
\\ dA(t)^{\#} \end{array}\right];\, X(0)=x. \nonumber\\
dY(t)&= C X(t)dt+ DdA(t), \label{eq:qsde-out}
\end{align}
with
\begin{align*}
A&=2\Theta(R+\Im\{K^{\dag}K\});\; B=2i\Theta [\begin{array}{cc}
-K^{\dag}S & K^TS^{\#}\end{array}];\; \\
C&=K;\; D=S,
\end{align*}
where $A(t)=(A_1(t),\ldots,A_n(t))^T$ is a vector of input vacuum bosonic
noise fields and $Y(t)=(Y_1(t),\ldots,Y_n(t))^T$ is a vector of
{\em output fields} that results from the interaction of $A(t)$
with the harmonic oscillator. Note that the dynamics of $X(t)$ and
$Y(t)$ are linear. We refer to $(A,B,C,D)$ as the {\em system
matrices} of $G$. For the case when $n=1$, we shall often refer to
the linear quantum stochastic system as a one degree of freedom
(open quantum harmonic) oscillator.

\begin{figure}[h!]
\centering
\includegraphics[scale=0.5]{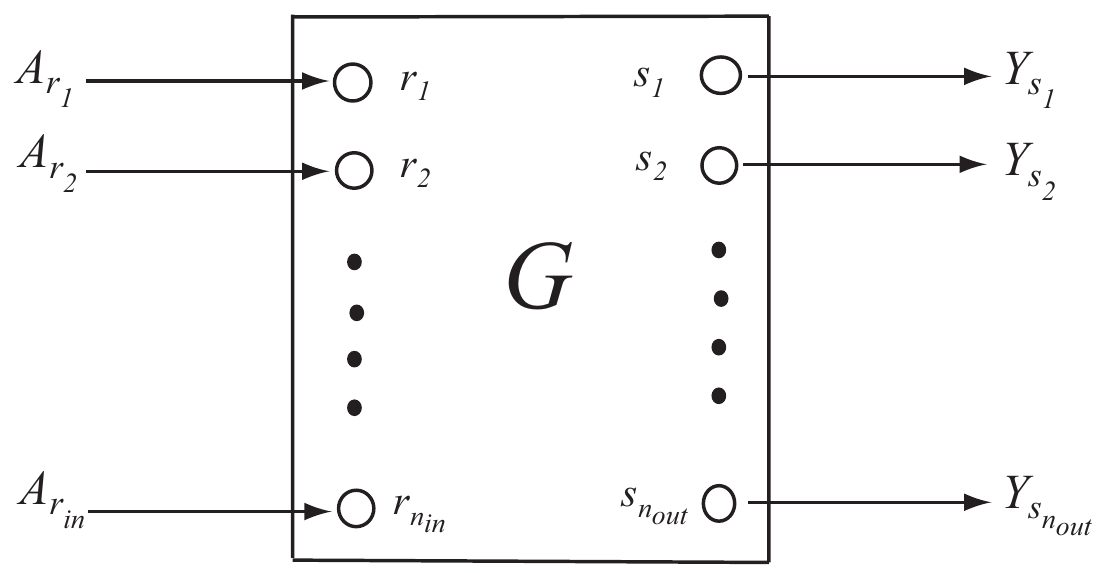}
\caption{A linear quantum stochastic system viewed as a device with $n_{\rm in}$ input ports
and $n_{\rm out}$ output ports.} \label{fig:Gdevice}
\end{figure}

Elements of $A(t)$ and $Y(t)$ may be partitioned into blocks. For
example, $A(t)$ may be partitioned as
$A(t)=(A_{r_1}(t)^T,\ldots,A_{r_{n_{\rm in}}}(t)^T)$ and $Y(t)$ as
$Y(t)=(Y_{s_1}(t)^T,\ldots,Y_{s_{n_{\rm out}}}(t)^T)^T$, where
$A_{r_j}(t)$ and $Y_{s_k}(t)$ are vectors of bosonic input and
output field operators of length $n_{r_j}$ and $n_{s_k}$,
respectively, such that $\sum_{j=1}^{n_{\rm in}} n_{r_j}=
\sum_{k=1}^{n_{\rm out}}n_{s_k}=m$. We refer to $n_{r_j}$ and
$n_{s_k}$ as the {\em multiplicity} of $A_{r_j}(t)$ and
$Y_{s_k}(t)$, respectively. It is important to keep in mind that
the sum of the multiplicities of all input and output partitions
sum up to $m$, the total number of all input and output fields.
With this partitioning, a linear quantum stochastic system $G$ may
be viewed as a quantum device having $n_{\rm in}$ {\em input
ports} and $n_{\rm out}$ {\em output ports} as illustrated in
Figure~\ref{fig:Gdevice}. The {\em multiplicity of a port} is then
defined as the multiplicity of the input or output fields coming
into or going out of that port.

Let us also recall the notion of the concatenation ($\boxplus$) and
series products  ($\triangleleft$) between open quantum systems
developed in \cite{GJ07}. The concatenation product $G_1 \boxplus
G_2$ between $G_1=(S_1,L_1,H_1)$ and $G_2=(S_2,L_2,H_2)$ defines
another quantum linear system given by:
$$
G_1 \boxplus G_2=(\diag(S_1,S_2),[\begin{array}{cc} L_1^T & L_2^T
\end{array}]^T,H_1+H_2),
$$
while if $G_1$ and $G_2$ have the same number of input (and
output) fields the series product $G_2 \triangleleft G_1$ defines
another linear quantum system given by:
$$
G_2 \triangleleft G_1=(S_2S_1,L_2+S_2L_1
,H_1+H_2+\Im\{L_2^{\dag}S_2L_1\}).
$$
Note that in the definition of the concatenation and series
product, it is not required that the elements of $L_1$ and $L_2$
and of $H_1$ and $H_2$ are commuting. That is, $G_1$ and $G_2$
may possibly be sub-components of the same system. Also both
operations are associative,  but neither operations are commutative.
By the associativity, the operations
$\boxplus_{j=1}^n G_j=G_1 \boxplus G_2 \boxplus \ldots \boxplus
G_n$ and $G_{n} \triangleleft \ldots \triangleleft G_2 \triangleleft
G_1$ are unambiguously defined.

The concatenation product corresponds to collecting together the
parameters of $G_1$ and $G_2$ to form one larger concatenated
system, and the series product is a mathematical abstraction of
the physical operation of {\em cascading} $G_1$ onto $G_2$, that
is, passing the output fields of $G_1$ as the input fields to
$G_2$. The cascaded system is then another linear quantum
stochastic system with parameters given by the series product
formula.

\section{The model matrix and concatenation of model matrices}
\label{sec:model-matrix}
The system $G=(S,L,H)$ can also be represented using a so-called
model matrix \cite{GJ08}. This representation will be
particularly useful for the goals of the present paper. For $G$
the model matrix representation $M(G)$ is given by (the
partitioned matrix):
\begin{equation}
M(G)=\left[\begin{array}{cccc} -iH-\frac{1}{2}L^{\dag}L &
-L^{\dag}S
\\ L & S \end{array} \right],\label{eq:model-matrix}
\end{equation}
with the understanding that if $L$ is partitioned as
$L=(L_1^T,L_2^T,\ldots,L_{n_{\rm out}}^T)^T$ with $L_j \in m_j \times 2$ and $
\sum_{j=1}^{n_{\rm out}}m_j=m $ and $S$ is partitioned accordingly as
$S=[S_{jk}]_{j=1,\ldots,n_{\rm out},k=1,\ldots,n_{\rm in}}$ with $S_{jk} \in
\Cbb^{m_j \times m'_k}$ and $\sum_{k=1}^{n_{\rm in}}m'_k=m$, then the model
matrix above can be expressed with respect to this partition as:
%\begin{eqnarray}
%M(G)=\left[\begin{array}{cccc}
%-iH-\frac{1}{2}\sum_{j=1}^{n_{\rm out}}L_j^{\dag}L_j &
%-\sum_{j=1}^{n_{\rm out}}L_j^{\dag}S_{j1} & \ldots &  -\sum_{j=1}^{n_{\rm out}}L_j^{\dag}S_{jn_{\rm in}}\\
%L_1 & S_{11} & \ldots & S_{1n_{\rm in}} \\
%\vdots & \vdots & \ddots & \vdots \\
%L_{n_{\rm out}} & S_{n_{\rm out}1} & \ldots & S_{n_{\rm out}
%n_{\rm in}}
%\end{array} \right] \label{eq:model-partitioning}.
%\end{eqnarray}
\begin{eqnarray}
M(G)=\left[\begin{array}{cccc}
-iH-\frac{1}{2}\sum_{j=1}^{n_{\rm out}}L_j^{\dag}L_j &
-\sum_{j=1}^{n_{\rm out}}L_j^{\dag}S_{j1} & \ldots &  -\sum_{j=1}^{n_{\rm out}}L_j^{\dag}S_{jn_{\rm in}}\\
L_1 & S_{11} & \ldots & S_{1n_{\rm in}} \\
\vdots & \vdots & \ddots & \vdots \\
L_{n_{\rm out}} & S_{n_{\rm out}1} & \ldots & S_{n_{\rm out}
n_{\rm in}}
\end{array} \right] \label{eq:model-partitioning}.
\end{eqnarray}

Also, with respect to a particular partitioning of $M(G)$, it is
convenient to attach a {\em unique label} to each row and  column
of the partition. For example, for the partitioning
(\ref{eq:model-partitioning}) we may give the labels
$s_0,s_1,\ldots,s_{n_{\rm out}}$ for the first, second, ..., $n_{\rm out}+1$-th row of
$M(G)$, respectively, and $r_0,r_1,\ldots,r_{n_{\rm in}}$ for the first,
second, ..., $n_{\rm in}+1$-th column of $M(G)$, respectively. Moreover,
with respect to this labelling (and analogously for any other
labelling scheme chosen), elements of the blocks are denoted as:
%\begin{align*}
%M_{s_0 r_0}(G)&=-iH-\frac{1}{2}\sum_{j=1}^{n_{\rm out}}L_j^{\dag}L_j;\;\\
%M_{s_0 r_k}(G)
%&=-\sum_{j=1}^{n_{\rm out}}L_j^{\dag}S_{jk},\, k>0;\\
%M_{s_j r_0}(G)&=L_j,\,j>0;\;M_{s_j r_k}=S_{jk},\,j,k>0.
%\end{align*}
\begin{align*}
M_{s_0 r_0}(G)&=-iH-\frac{1}{2}\sum_{j=1}^{n_{\rm out}}L_j^{\dag}L_j;\;M_{s_0 r_k}(G)
=-\sum_{j=1}^{n_{\rm out}}L_j^{\dag}S_{jk},\, k>0;\\
M_{s_j r_0}(G)&=L_j,\,j>0;\;M_{s_j r_k}=S_{jk},\,j,k>0.
\end{align*}
Since $M(G)$ is another representation of a physical system
described by $G$, and $G$ can be identified directly from the
entries of $M(G)$, we will often omit the $G$ and for brevity
write a model matrix simply as $M$ and denote its entries by
$M_{\alpha \beta}$, with $\alpha$ ranging over row labels and
$\beta$ ranging over the column labels. Thus, we also refer to the
triple $S,L,H$ in (\ref{eq:model-matrix}) as {\em parameters of
the model matrix} $M$.

Several model matrices can be concatenated to form a larger model
matrix. Such a concatenation corresponds to collecting together
the model parameters of the individual matrices in a larger model
matrix and is again denoted by the symbol $\boxplus$. If
$G_1=(S_1,L_1,H_1)$ and $G_2=(S_2,L_2,H_2)$ then the concatenation
$M(G_1) \boxplus M(G_2)$ is defined as:
%\begin{align*}
%M(G_1) \boxplus M(G_2) &=\left[\begin{array}{ccc} -iH_1 -iH_2
%-\half L_1^{\dag}L_1 -\half L_2^{\dag}L_2 & -L_1^{\dag}S_1 &
%-L_2^{\dag}
%S_2\\
%L_1 & S_1 & 0 \\
%L_2 & 0  & S_2  \end{array} \right]\\
%&= M(G_1 \boxplus G_2).
%\end{align*}
\begin{align*}
M(G_1) \boxplus M(G_2) &=\left[\begin{array}{ccc} -iH_1 -iH_2
-\half L_1^{\dag}L_1 -\half L_2^{\dag}L_2 & -L_1^{\dag}S_1 &
-L_2^{\dag}
S_2\\
L_1 & S_1 & 0 \\
L_2 & 0  & S_2  \end{array} \right]\\
&= M(G_1 \boxplus G_2).
\end{align*}

\section{Edges, Elimination of Edges, and Reduced Markov Models}
\label{sec:edges}
Following \cite{GJ08}, a particular  row partition labelled $s_k$ with $k>0$
in a model matrix can be associated with an output port $s_k$
(having multiplicity $n_{s_k}$) while a particular  column partition
$r_j$ with $j>0$ can be associated with an input port $r_j$ (having
multiplicity $n_{r_j}$). In a system which is the concatenation of
several sub-systems, it is possible to
connect an output port $s_k$ from one sub-system to an input
port $r_j$ of another sub-system (possibly the same sub-system to
which the output port belongs) to form what is called
an {\em internal edge} denoted by $(s_k,r_j)$. For this connection
to be possible, the ports $s_k$ and $r_j$ {\em must
have the same multiplicity}. Such an edge then represents a
{\em channel} from port $s_k$ to port $r_j$. All ports which are
connected to other ports to form an internal edge or channel are referred
to as {\em internal ports} and fields coming into or leaving such ports are called
{\em internal fields}. All other input and output ports that
are not connected in this way are viewed as having semi-infinite
edges (since they do not terminate at some input or output
port, as appropriate) and are referred to as {\em external ports}
and the associated semi-infinite edges are referred to as {\em
external edges}. Fields coming into or leaving external ports are called
{\em external fields}. From a point of view in line with circuit theory,
one may think of a linear quantum stochastic system as being a ``node'' on a network and
quantum fields as quantum ``wires'' that can connect different nodes.

In any internal edge $(s_k,r_j)$, there is a finite delay present
due to the time which is required for the signal from port $s_k$
to travel to  port $r_j$. As a consequence of these finite time
delays, concatenated systems with internal edges cannot be
represented by a Markov model such as presented in Section
\ref{sec:linear-summary} (see \cite{NJD08} and the references
therein for an overview of Markov models). However, as shown in
\cite{GJ08}, the non-Markov model converges to a {\em reduced}
Markov model in the limit that the time delay on all internal
edges go to zero. That is, for negligibly small time delays, the
reduced Markov model acts as an approximation of the non-Markov
model. In particular, such a reduced model serves as a powerful
approximation of quantum optical networks in which signals travel
at the speed of light and the time delay can be considered to be
practically zero if the internal input and output ports are not
extremely far apart. %As the time delay in every internal edge is
%allowed to go to zero in the limit, what happens is that the edges
%become  eliminated, the associated input and output ports
%associated with the edges merge, and the associated input and
%output fields are removed from the concatenated model. This
%results in a simpler approximate Markov model with a fewer number of
%input and output fields.
We recall the following results:

\begin{theorem}
\label{thm:qfn-red} {\bf \cite[Theorem 12 and Lemma 16]{GJ08}} Let
$\tau_{(s_k,r_j)}$, $j,k>0$, be the time delay for an internal
edge $(s_k,r_j)$ and assume that $I-S_{k j}$ is invertible. Then
in the limit that $\tau_{(s_k,r_j)} \downarrow 0 $, $M(G)$ with
the edge $(s_k,r_j)$ connected reduces to a simplified model
matrix $M_{\rm red}$ with input ports labelled
$r_0,r_1,\ldots,r_{j-1},r_{j+1},\ldots,r_{n_{\rm out}}$ and output
ports labelled $s_0,s_1,\ldots,s_{k-1},s_{k+1},\ldots,s_{n_{\rm
in}}$ (i.e., the connected ports $r_j$ and $s_k$ are removed from
the labelling and the associated row and column removed from
$M(G)$). The block entries of $M_{\rm red}$ are given by:
\begin{align*}
(M_{\rm red})_{\alpha \beta}=M_{\alpha \beta} + M_{\alpha r_j}(1-S_{kj})^{-1}M_{s_k \beta},
\end{align*}
with $\alpha \in \{s_0,s_1,\ldots,s_{n_{\rm out}}\}\backslash
\{s_k\}$ and $\beta \in \{r_0,r_1,\ldots,r_{n_{\rm
in}}\}\backslash \{r_j\}$. $M_{\rm red}$ is the model matrix of a
linear quantum stochastic system $G_{\rm red}$ with parameters:
\begin{align*}
(S_{\rm red})_{pq}&=S_{pq} +S_{pj}(I-S_{kj})^{-1}S_{kq} \\
(L_{\rm red})_{p}&= L_{p} + S_{pj} (I-S_{kj})^{-1}L_{k}\\
H_{\rm red}&= H + \sum_{p=1}^{n_{\rm out}} \Im\{L_{p}^{\dag}S_{pj}(I-S_{kj})^{-1}L_{p}\},
\end{align*}
for all $p \in \{1,2,\ldots,n_{\rm out}\}\backslash \{k\}$ and $q \in  \{1,2,\ldots,n_{\rm in}\}\backslash \{j\}$.
\end{theorem}

Several internal edges may  be eliminated one at a time in a
sequence leading to a corresponding sequence of reduced model
matrices. The following result shows that such a procedure is unambiguous:
%It turns out, rather conveniently, that the particular
%order with which a collection of edges are eliminated is
%irrelevant and the resulting final reduced model matrix (after
%elimination of the final edge) is unambiguously determined:
\begin{theorem}
{\bf \cite[Lemma 17]{GJ08}} The reduced model matrix obtained after eliminating all the
internal edges in a set of internal edges one at a time is
independent of the sequence in which these edges are eliminated.
\end{theorem}

Suppose that $M$ can be partitioned as:
%\begin{eqnarray}
%\left[\begin{array}{ccc} -iH-1/2L_{\rm i}^{\dag}L_{\rm i}-1/2L_{\rm e}^{\dag}L_{\rm e} &
%-L_{\rm i}^{\dag}S_{\rm ii}-L_{\rm i}^{\dag}S_{\rm ei} & -L_{\rm e}^{\dag}S_{\rm ie}-L_{\rm i}^{\dag}S_{\rm ii}\\
%L_{\rm i} & S_{\rm ii} & S_{\rm ie} \\
%L_{\rm e} & S_{\rm ei} & S_{\rm ee}
%\end{array}
%\right], \label{eq:in-out-partition}
%\end{eqnarray}
\begin{eqnarray}
\left[\begin{array}{ccc} -iH-1/2L_{\rm i}^{\dag}L_{\rm i}-1/2L_{\rm e}^{\dag}L_{\rm e} &
-L_{\rm i}^{\dag}S_{\rm ii}-L_{\rm i}^{\dag}S_{\rm ei} & -L_{\rm e}^{\dag}S_{\rm ie}-L_{\rm i}^{\dag}S_{\rm ii}\\
L_{\rm i} & S_{\rm ii} & S_{\rm ie} \\
L_{\rm e} & S_{\rm ei} & S_{\rm ee}
\end{array}
\right], \label{eq:in-out-partition}
\end{eqnarray}
where the subscript ${\rm i}$ refers to ``internal'' and ${\rm e}$
to ``external''. That is, parameters with subscript ${\rm i}$ or
${\rm ii}$ pertain to internal ports, those with subscript ${\rm
e}$ or ${\rm ee}$ pertain to external ports, while $S_{\rm ie}$
and $S_{\rm ei}$ pertain to scattering of internal fields to
external fields and vice-versa, respectively. Interconnection
among internal input and output ports can then be conveniently
encoded by a so-called {\em adjacency matrix}. Let $n_{\rm i}$
denote the total multiplicity of internal input and output ports
and let us view a port with multiplicity $k$ as $k$ distinct ports
of multiplicity $1$. Suppose that these multiplicity 1 ports are
numbered consecutively starting from 1, then an adjacency matrix
$\eta$ is  an $n_{\rm i} \times n_{\rm i}$ square matrix whose
entries are either $1$ or $0$ with $\eta(j,k)=1$ ($j,k \in
\{1,2,\ldots,n_{\rm i}\})$ only if the $j$-th output port and the
$k$-th input port form a channel or internal edge. Note that at
most only a single element in any row or column of $\eta$ can take
the value 1. Internal edges can be simultaneously eliminated as
follows:
%We have:
\begin{theorem}
\label{thm:qfn-simul-elim}{\bf \cite[Section 5]{GJ08}} Suppose that $M$ has a partitioning
based on internal and external components as in
(\ref{eq:in-out-partition}) and that
 connections between internal ports have been encoded in an adjacency matrix $\eta$. If $(\eta-S_{\rm ii})^{-1}$ exists, then the reduced model
matrix $M_{\rm red}$ after simultaneous elimination of all internal edges has the parameters:
%\begin{eqnarray*}
%\left[\begin{array}{cc} -iH_{\rm red}-1/2L_{\rm red}^{\dag}L_{\rm red} &
%-L_{\rm red}^{\dag}S_{\rm red} \\
%L_{\rm red} & S_{\rm red}
%\end{array}\right],
%\end{eqnarray*}
%with
\begin{align*}
S_{\rm red}&=S_{\rm ee}+S_{\rm ei}(\eta-S_{\rm ii})^{-1}S_{\rm ie}\\
L_{\rm red}&=L_{\rm e}+S_{\rm ei} (\eta-S_{\rm ii})^{-1}L_{\rm i}\\
H_{\rm red}&=H+\sum_{j={\rm i,e}} \Im\{L_j^{\dag}S_{j {\rm
i}}(\eta-S_{\rm ii})^{-1}L_{\rm i}\}.
\end{align*}
\end{theorem}
%\begin{remark}
%It is important to keep in mind that a reduced model matrix is an
%approximation to the actual non-Markov system that results from
%the formation of internal edges and channels when the delays in
%propagating through.
%\end{remark}

\section{Prior work}
\label{sec:prior} Let $G=(I_m,Kx,\half x^T R x)$ be a linear
quantum stochastic system with $K=[\begin{array}{ccc} K_1 & \ldots
& K_n
\end{array}] \in \Cbb^{m \times 2n}$ and $R=[R_{jk}]_{j,k=1,\ldots,n} \in \Rbb^{2n \times 2n}$ ($R_{kj}=R_{jk}^T$).
Let $x_j=(q_j,p_j)^T$ so that $x=(x_1^T,\ldots,x_n^T)^T$ and $x$ satisfies the commutation relations of Section
\ref{sec:linear-summary}. Then the
following result holds:
\begin{theorem}
\cite{NJD08} Let $G_k=(I,K_k x_k,\half x_k^T R_{kk}x_k)$.
Then \[ G=(0,0,H^d) \boxplus (G_n \triangleleft \ldots
\triangleleft G_{2} \triangleleft G_1),
\]
where $H^d=\sum_{j=1}^{n-1} \sum_{k=j+1}^{n} H^d_{jk}$ and $H^d_{jk}=x_k^T (R_{jk}^T-\Im\{K_k^{\dag}K_j\})x_j$.
\end{theorem}

\begin{figure}[h!]
\centering
\includegraphics[scale=0.5]{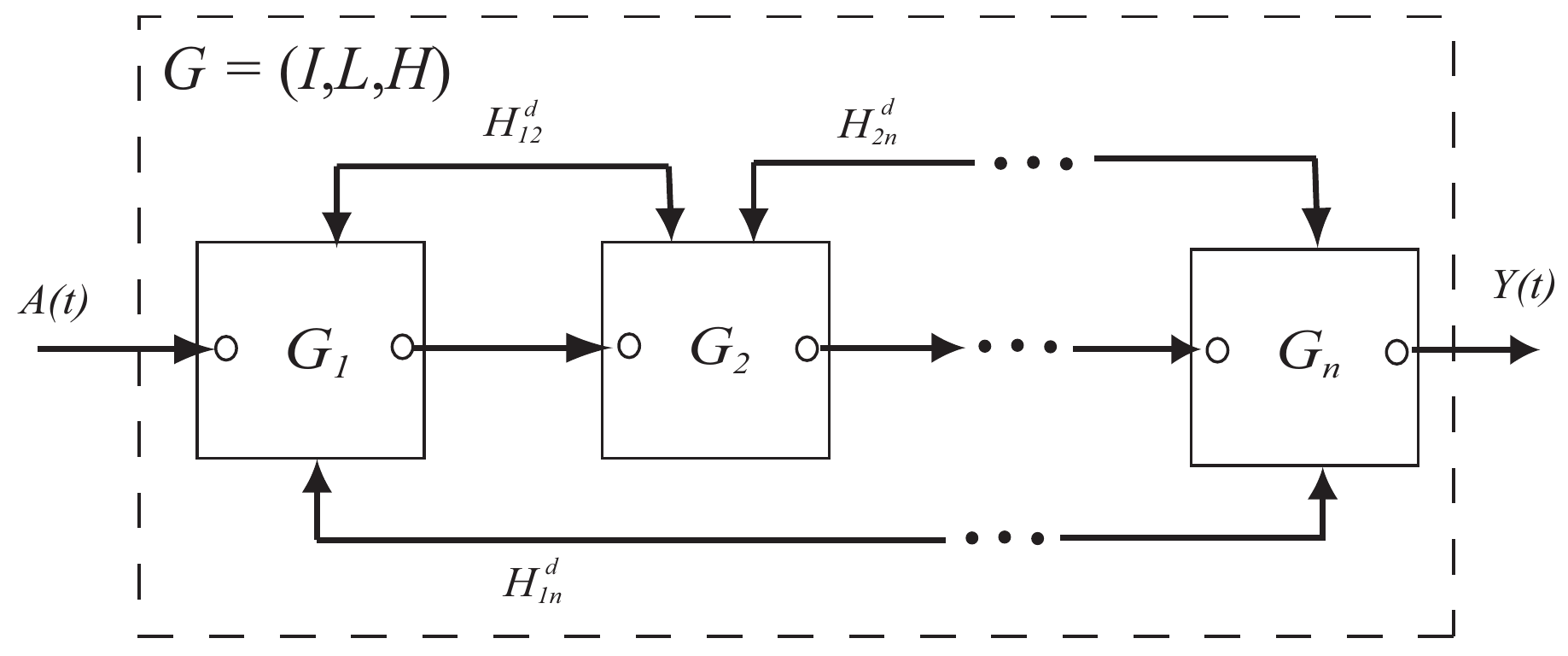}
\caption{Realization of $G=(I,L,H)$ via a cascade connection
of one degree of freedom quantum harmonic oscillators $G_j$ and bilinear
direct interaction Hamiltonians $H_{jk}$ between every pair of oscillators}
\label{fig:njd-cascade}
\end{figure}

The theorem says that $G$ can in principle be constructed as the
cascade connection $G_n \triangleleft \ldots \triangleleft G_{2}
\triangleleft G_1$ of the one degree of freedom oscillators
$G_1,G_2,\ldots,G_n$, together with a direct bilinear interaction
Hamiltonian $H^d$ which is the sum of bilinear interaction
Hamiltonians $H^d_{jk}$ between each pair of oscillator $G_j$ and
$G_k$. This construction is depicted in
Figure~\ref{fig:njd-cascade}. It was then shown how each $G_k$ can
be constructed from certain basic quantum optical components and
how $H^d_{jk}$ can be implemented between any pair of oscillators.
However, a drawback of this approach, based on what is feasible
with current technology, is the challenging nature of implementing
$H^d$. This may possibly involve complex positioning and
orientation of the oscillators and thus practically challenging
for systems with more than just a few degrees of freedom. Although
advances in experimental methods and emergence of new technologies
may eventually alleviate this difficulty, it is naturally of
immediate interest to also explore alternative methods of
implementing this interaction Hamiltonian, at least approximately.
In the next section, we propose such an alternative synthesis by
exploiting the theory of quantum feedback networks that has been
elaborated upon in preceding sections of the paper.

\section{Main synthesis results}
\label{sec:main-results} For $j=1,\ldots,n$, let
$G_{jk}=(S_{jk},L_{jk},0)$ for $k=1,\ldots,n,k \neq j$, and
$G_{jj}=(S_{jj},L_j,H_j)$, with $S_{jk} \in \Cbb^{c_{jk} \times
c_{jk}}$, $c_{jk}=c_{kj}$ and $c_{jj}=m$, $L_{jk}=K_{jk}x_j$,
and $H_j=\half x_j^T R_j x_j$ with $R_{j}=R_j^T \in \Rbb^{2 \times 2}$.
Here $x_j=(q_j,p_j)^T$ is as defined in the previous section. Let $G_{j}=\boxplus_{k=1}^{n} G_{jk}$
for $j=1,\ldots,n$, and note that $G_j=(S_j,L_j,H_j)$ with
$S_j=\diag(S_{j1},S_{j2},\ldots,S_{jn})$, $L_j =
(L_{j1}^T,L_{j2}^T,\ldots,L_{jn}^T)^T$, and $H_j$ as already defined.

\begin{figure}%[h!]
\centering
\includegraphics[scale=0.55]{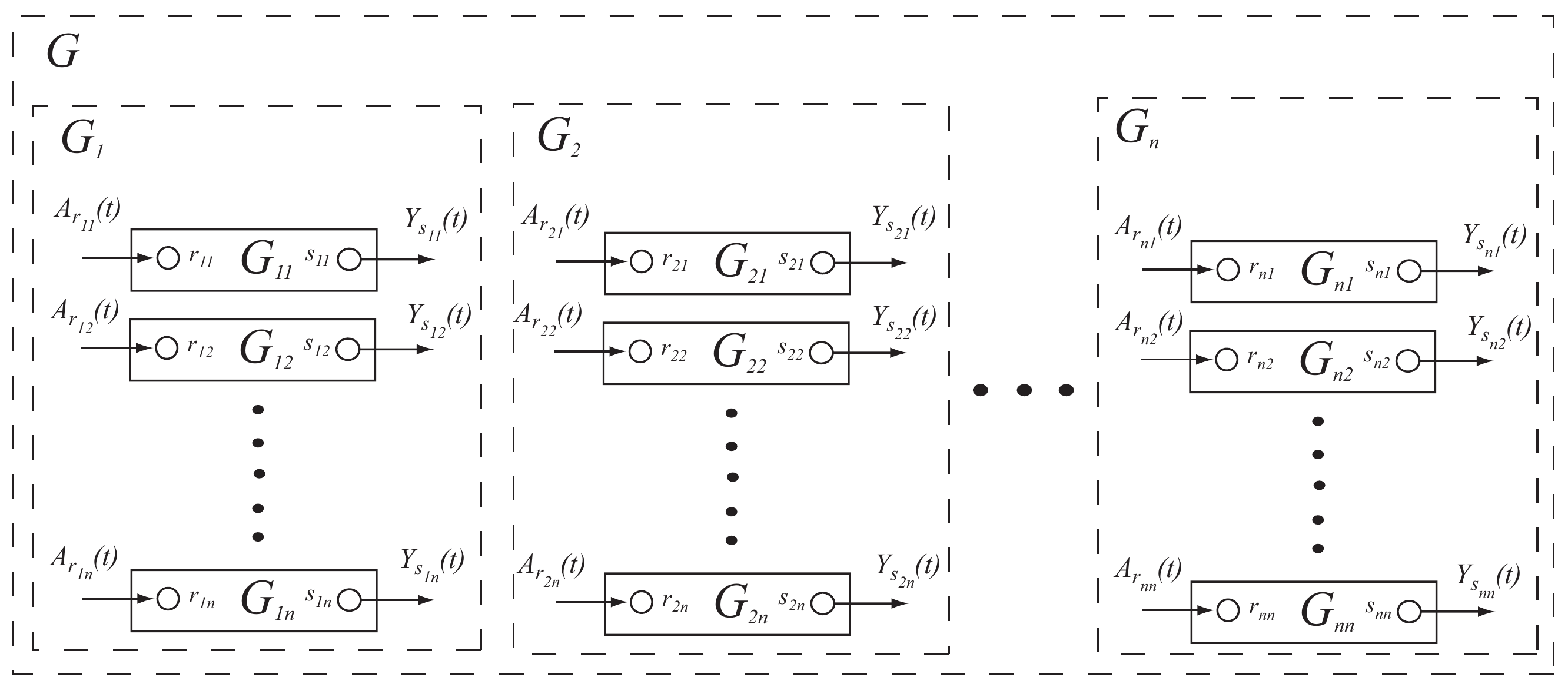}
\caption{The concatenation decomposition of $G$ as
$G=\boxplus_{j=1}^n \boxplus_{k=1}^n G_{jk}$.}
\label{fig:G-concat-decom}
\end{figure}

Consider now the model matrix $M$ for the concatenated system
$\boxplus_{j=1}^n G_j$, see Figure~\ref{fig:G-concat-decom}. With
respect to the natural partitioning of $M$ induced by the  $G_{jk}$'s
(via their concatenation), we label the first $n+1$ rows of $M$ as
$s_{00},s_{11},\ldots,s_{1n}$, the next $n$ rows as
$s_{21},\ldots,s_{2n}$, the $n$ rows after that
$s_{31},s_{32},s_{33},\ldots,s_{3n}$ and so on until the last $n$
rows are labelled $s_{n1},s_{n2},\ldots,s_{nn}$. Similarly, we
label the first $n+1$ columns of $M$ as
$r_{00},r_{11},\ldots,r_{1n}$, the next $n$ columns as
$r_{21},\ldots,r_{2n}$, the $n$ columns after that
$r_{31},\ldots,r_{3n}$ and so on until the last $n$ columns
$r_{n1},r_{n2}\ldots,r_{nn}$. On occasions, we will need to write
a bracket around one of both of the subscripts of $r$ or $s$
(e.g., as in $s_{(n-1)k}$ or $r_{(k+1)(k+1)}$).

\begin{theorem}
\label{thm:qfn-construct} Let the output port $s_{jk}$ be
connected to the input port $r_{kj}$ to form an internal
edge/channel $e_{jk}=(s_{jk},r_{kj})$ for all $j,k=1,\ldots,n,j\neq k$.
Assuming that $\left[\begin{array}{cc} -S_{jk} & I
\\ I & -S_{kj}\end{array} \right]$ is invertible $\forall
j,k=1,\ldots,n,j\neq k$, then the reduced model matrix
$M_{\rm red}$ obtained by allowing the delays in all internal
edges $\{e_{jk},\; j \neq k\}$ go to zero has parameters given by:
\begin{eqnarray*}
S_{\rm red} &=& \diag(S_{11},S_{22},\ldots,S_{nn})\\
L_{\rm red}&=& (L_{11}^T,L_{22}^T,\ldots,L_{nn}^T)^T\\
H_{\rm red} &=&\sum_{k=1}^n H_k + \sum_{j=1}^{n-1} \sum_{k=j+1}^n
\Im\biggl\{[\begin{array}{cc} L_{jk}^{\dag} &
L_{kj}^{\dag}\end{array}] \left[\begin{array}{cc} I & -S_{jk} \\
-S_{kj} & I \end{array} \right]^{-1}\left[\begin{array}{c}
L_{jk} \\ L_{kj} \end{array} \right]\biggr\}.
\end{eqnarray*}
\end{theorem}

The proof of the theorem is given in Appendix
\ref{sec:proof-qfn-construct}. We will also exploit the following lemma:
\begin{lemma}
\label{lem:direct-coupling} For any real $2 \times 2$ matrix $R$ and unitary complex numbers
$S_{12}$ and $S_{21}$ satisfying $S_{12}S_{21} \neq 1$, there
exist $1 \times 2$ complex matrices $K_1$ and $K_2$ such that
$R-\Im\{\frac{S_{12}}{1-S_{12}S_{21}} K_1^{\dag}K_2+ \frac{S_{21}}{1-S_{12}S_{21}} K_1^T K_2^{\#}\}=0$. In fact,
a pair $K_{1}$ and $K_{2}$ satisfying this is given by
$K_1=[\begin{array}{cc} \kappa & i\kappa  \end{array}]$ with
$\kappa$ an arbitrary non-zero real number and
$K_2=2i[\begin{array}{cc} 1 & 0 \end{array}][\begin{array}{cc}
-K_1^{\dag}\Delta^* & K_1^T \Delta \end{array}]^{-1}R$, where
$\Delta=2\frac{S_{21}-S_{12}^*}{|1-S_{12}S_{21}|^2}$. Or, alternatively,
$K_2=[\begin{array}{cc} \kappa & i\kappa
\end{array}]$ and \\
$K_1=2i[\begin{array}{cc} 1 & 0
\end{array}][\begin{array}{cc} K_2^{\dag}\Delta & -K_2^T
\Delta^*\end{array}]^{-1}R^T$.
\end{lemma}

See Appendix \ref{sec:proof-direct-coupling} for a proof of the lemma.
As a consequence of the above theorem and lemma, we have the
following result:
\begin{corollary}
\label{cor:qfn-synthesis} Let $c_{jk}=c_{kj}=1$ whenever $j \neq
k$, and $c_{jj}=m$ for all $j,k=1,\ldots,n$. Also, let
$S_{jj}=I_{m}$, $S_{kj}=e^{i\theta_{kj}}$ and
$S_{jk}=e^{i\theta_{jk}}$ with $\theta_{kj}, \theta_{jk} \in [0,2\pi)$
satisfying $\theta_{kj}+\theta_{jk}\neq 0$, $K_{jj}=K_j$ and
$R_j=R_{jj}-2{\rm sym}\biggl(\sum_{k=1,k\neq j}^n \Im\bigl\{\frac{1}{1-S_{jk}S_{kj}}K_{jk}^{\dag}K_{jk}\bigr\}\biggr)$ ($K_{j}$
and $R_{jj}$ given), where ${\rm sym }(A)=\half (A+A^T)$, and the pair $(K_{jk},K_{kj})$ ($j \neq k$) be given by:
\begin{eqnarray}
K_{jk}&=&[\begin{array}{cc} \kappa_{jk} & i\kappa_{jk} \end{array}] \nonumber \\
K_{kj}&=&2i[\begin{array}{cc} 1 & 0 \end{array}][\begin{array}{cc}
 -K_{jk}^{\dag}\Delta_{jk}^* & K_{jk}^T\Delta_{jk}\end{array}]^{-1}(R_{jk}-\Im\{K_j^T K_k^{\#}\}),
\label{eq:K-coupling-1}
\end{eqnarray}
or
\begin{eqnarray}
K_{kj}&=&[\begin{array}{cc} \kappa_{jk} & i\kappa_{jk} \end{array}] \nonumber \\
K_{jk}&=&2i[\begin{array}{cc} 1 & 0 \end{array}][\begin{array}{cc}
K_{kj}^{\dag}\Delta_{jk} & -K_{kj}^T\Delta_{jk}^*\end{array}]^{-1}(R_{jk}-\Im\{K_j^T K_k^{\#}\})^T, \label{eq:K-coupling-2}
\end{eqnarray}
where $R_{jk}=R_{kj}^T \in \Rbb^{2 \times 2}$,
$\Delta_{jk}=2\frac{S_{kj}-S_{jk}^*}{|1-S_{kj}S_{jk}|^2}$ and
$\kappa_{jk}$ is an arbitrary non-zero real constant for all
$j,k$. Then the reduced Markov model $G_{\rm red}=(S_{\rm
red},L_{\rm red},H_{\rm red})$ has the decomposition $G_{\rm
red}=\boxplus_{k=0}^{n} G_{\rm red,k}$ with $G_{\rm red,
0}=(0,0,H_{\rm red})$ and $G_{\rm red,k}=(S_{kk},L_{kk},0)$ for
$k=1,\ldots,n$. Moreover, the network $G_{\rm net}=(S_{\rm
net},L_{\rm net},H_{\rm net})$ formed by forming the series
product of $G_{\rm red,n} \triangleleft \ldots \triangleleft  
G_{{\rm red},2} \triangleleft G_{\rm red,1}$ within the
concatenated system $G_{\rm red}$ and defined by $G_{\rm
net}=G_{\rm red,0} \boxplus (G_{\rm red,n} \triangleleft \ldots \triangleleft  G_{{\rm red},2} \triangleleft G_{\rm red,1})$ is a
linear quantum stochastic system with parameters given by:
\begin{eqnarray*}
S_{\rm net}&=&I_{m};\\
L_{\rm red}&=& Kx,\;K=[\begin{array}{cccc} K_1
& K_2 & \ldots & K_n \end{array}];\\
H_{\rm red}&=& \half x^T R x,\; R=[R_{jk}]_{j,k=1,\ldots,n}.
\end{eqnarray*}
In other words, $G_{\rm net}$ realizes a linear quantum stochastic
system with the above parameters.
\end{corollary}

\begin{remark}
\label{rem:cascade-as-reduced} Note that the series connection
$G_{\rm red,n} \triangleleft \ldots \triangleleft  G_{{\rm red},2} \triangleleft G_{\rm
red,1}$ can be viewed as forming and futher eliminating the
internal edges $\{(s_{kk},r_{(k+1)(k+1)});\,k=1,\ldots,n-1\}$ in
$G_{\rm red}$ or $G$ and, hence, is in essence a special case of a
reduced Markov model \cite{GJ08}.
\end{remark}

The proof of the corollary is given in Appendix \ref{sec:proof-qfn-synthesis}.
The corollary shows that an arbitrary linear quantum stochastic
system $(I,Kx,\half x^T R x)$ can be realized by a quantum network
$G_{\rm net}$ constructed according to Theorem
\ref{thm:qfn-construct} and the corollary, with an appropriate
choice of the parameters $R_{j}$, $K_{jk}$ and $S_{kj}$ ($j \neq k$). From here, any system $(S,Kx,\half x^T R x)$
can then be easily obtained as $(S,Kx,\half x^T R x)=(I,Kx,\half x^T R x) \triangleleft (S,0,0)$. That
is, as a cascade of a static network that realizes the unitary
scattering matrix $S$ and the system $(I,Kx,\half x^T R x)$ (see \cite{GJ07,NJD08}).
\begin{example}
Consider the realization problem of a two degrees of freedom linear
quantum stochastic system $G_{\rm sys}=(I,Kx,1/2 x^T R x)$
($x=(q_1,p_1,q_2,p_2)^T$) with
\begin{align*}
K &=[\begin{array}{cc} K_1 & K_2
\end{array}]=[\begin{array}{cccc} 3/2 & i/2 & 1 & i \end{array}]\;  (K_j \in \Cbb^{1 \times 2},\;j=1,2)
;\\
R &=\left[\begin{array}{cc} R_{11}& R_{12} \\ R_{12}^T & R_{22}
\end{array} \right]=\left[\begin{array}{cccc} 2 & 0.5 & 1 & 1 \\
0.5 & 3 & -1 & -1\\
1 & -1 & 1 & 0 \\
1 & -1 & 0 & 1 \end{array} \right]\; (R_{jk} \in \Rbb^{2 \times
2},\;j,k=1,2).
\end{align*}
Let $x_1=(q_1,p_1)^T$ and $x_2=(q_2,p_2)^T$. Define
$G_1=\biggl(\diag(I_m,S_{12}),\left[\begin{array}{c} K_{11} \\ K_{12}
\end{array}\right]x_1,\half x_1^T R_{1} x_1\biggr)$ and
$G_2=\biggl(\diag(S_{21},I_m),\left[\begin{array}{c} K_{21} \\ K_{22}
\end{array}\right]x_2,\half x_2^T R_{2} x_2\biggr)$, with certain parameters still to be determined.
Choose $\theta_{12}=0$ and $\theta_{21}=\pi/2$, so that $\theta_{12}+\theta_{21}
\neq 0$ as required in Corollary \ref{cor:qfn-synthesis}. Then set
$S_{12}=e^{i\theta_{12}}=1$, $S_{21}=e^{i\theta_{21}}=i$, $K_{11}=K_1=[\begin{array}{cc}
3/2 & 1/2 \end{array}]$ and $K_{22}=K_2=[\begin{array}{cc} 1 & i
\end{array}]$. Compute
$\Delta_{12}=2\frac{S_{21}-S_{12}^*}{|1-S_{12}S_{21}|^2}=-1+i$ and
set $\kappa_{12}=1$. Then by Corollary \ref{cor:qfn-synthesis} we
set $K_{12}=[\begin{array}{cc} \kappa_{12} & i\kappa_{12}
\end{array}]=[\begin{array}{cc} 1 & i \end{array}]$ and compute
$K_{21}=[\begin{array}{cc} 1.25-0.25i & 1.75+0.75i \end{array}]$,
$R_1=\left[\begin{array}{cc} 1 & 0.5 \\ 0.5 &
2\end{array}\right]$ and $R_2=-\left[\begin{array}{cc} 0.625 & 2 \\
2 & 2.625 \end{array}\right]$. Thus, we have determined all the
parameters of $G_1$ and $G_2$. Labelling the ports of $G_1$ and
$G_2$ according to the convention adopted in this section, $G_{\rm
sys}$ can be implemented by concatenating $G_1$ and $G_2$ and
eliminating the internal edges $(s_{12},r_{21})$ and
$(s_{21},r_{12})$ to form $G_{\rm red}$ and then eliminating the
edge $(s_{11},r_{22})$ (cf. Remark \ref{rem:cascade-as-reduced})
to obtain $G_{\rm net}$ as an approximate realization of $G_{\rm
sys}$. This realization is illustrated in
Figure~\ref{fig:Gsys-realization}. $G_1$ and $G_2$ can then  be
physical realized in the quantum optics domain following the
constructions proposed in \cite{NJD08}. Using the schematic
symbols of \cite{NJD08}, a quantum optical circuit that is a
physical realization of $G_{\rm sys}$ is depicted in
Figure~\ref{fig:Gsys-circuit}.
\begin{figure}[h!]
\centering
\includegraphics[scale=0.45]{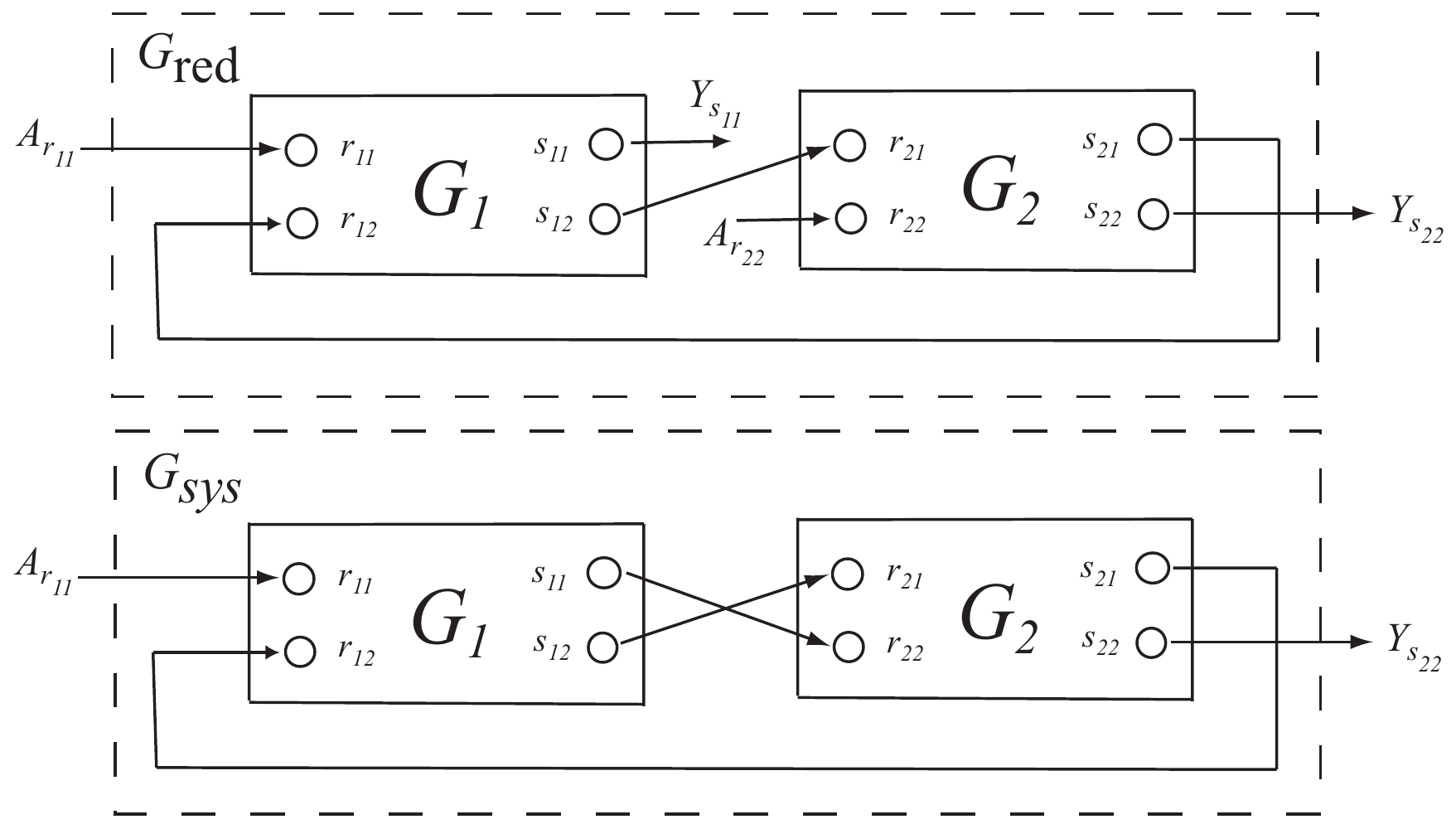}
\caption{Realization of $G_{\rm sys}$ via a quantum feedback network. In the top figure the internal edges
$(s_{12},r_{21})$ and $(s_{21},r_{12})$ are formed and eliminated to obtain a reduced Markov model $G_{\rm red}$.
Then in the bottom figure a series (cascade) connection is formed by the eliminating the internal edge
$(s_{11},r_{22})$ to realize $G_{\rm sys}$}
\label{fig:Gsys-realization}
\end{figure}

\begin{figure}%[h!]
\centering
\includegraphics[scale=0.5]{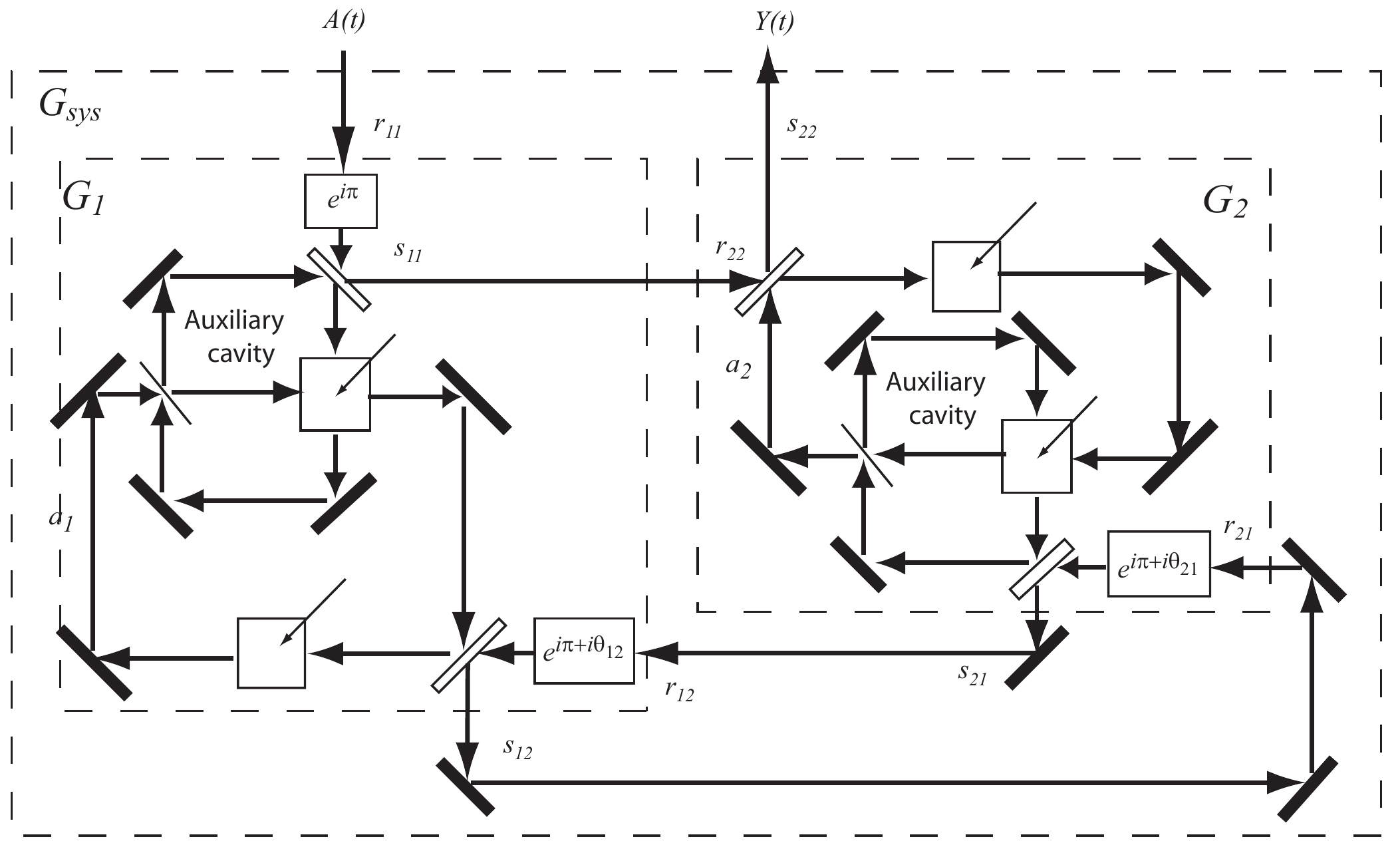}
\caption{A quantum optical circuit that realizes $G_{\rm sys}$
according to the quantum feedback network in
Figure~\ref{fig:Gsys-realization}; the schematic symbols are taken
from \cite{NJD08}. Here $a_j=\frac{q_j+ip_j}{2}$ is cavity mode of
the optical cavity around which the physical realization of the
oscillator $G_j$ is based, $j=1,2$. The dashed box labelled $G_j$
is the part of the circuit realizing $G_j$, and the input and
output ports of $G_j$ are indicated by their respective labels.
The value of the parameters of devices in this circuit can be
determined according to \cite{NJD08}.} \label{fig:Gsys-circuit}
\end{figure}

\end{example}

\section{Synthesis of passive systems}
\label{sec:passive}
Let us now consider a special class of linear quantum stochastic systems
that we shall refer to as {\em passive} linear quantum stochastic systems, for a reason
that is explained below, and show that any such system can be built
from passive components. This class of systems has also been considered in, e.g., \cite{GGY08,MP09}.

For $k=1,\ldots,n$ let $a_k=\frac{q_k+ip_k}{2}$ be the
annihilation operators for mode $k$ and define
$a=(a_1,\ldots,a_n)^T$. Then $a$ satisfies the CCR
\begin{eqnarray*}
&&\left[\begin{array}{c} a \\
a^{\#}\end{array}\right]\left[
\begin{array}{cc} a^{\dag} & a^T \end{array}\right]-\left(\left[\begin{array}{c} a^{\#} \\ a \end{array}\right]\left[
\begin{array}{cc} a^T & a^{\dag} \end{array}\right]\right)^T\\
&&\quad = \diag(I_{n},-I_{n}).
\end{eqnarray*}
Moreover, note that:
$$
\left[\begin{array}{c} a \\
a^{\#}\end{array}\right]=\left[\begin{array}{c} \Sigma \\
\Sigma^{\#} \end{array} \right]x;\quad
$$
where
$$
\Sigma =\left[\begin{array}{ccccccc} \half & i\half & 0 & 0 & 0 &
\ldots & 0\\
0 & 0 & \half & i\half & 0 &  \ldots & 0\\
\vdots & \ddots & \ddots & \ddots & \ddots & \ddots & \vdots \\
0 & \ldots & \ldots & \ldots & 0 & \half & i\half
\end{array}\right].
$$
We also make note that $\left[\begin{array}{c} \Sigma \\
\Sigma^{\#} \end{array} \right]^{-1}=2\left[\begin{array}{cc}
\Sigma^{\dag} & \Sigma^{T} \end{array} \right]$ and therefore we have
$$
x= \left[\begin{array}{c} \Sigma \\
\Sigma^{\#} \end{array} \right]^{-1}\left[\begin{array}{c} a \\
a^{\#}\end{array}\right]=2\left[\begin{array}{cc} \Sigma^{\dag} &
\Sigma^{T} \end{array} \right]\left[\begin{array}{c} a \\
a^{\#}\end{array}\right].
$$

An $n$ degrees of freedom system $G=(S,Kx,\half x^T R x)$ is said to be {\em passive} if
we may write $H=\half x^T Rx = \half a^{\dag} \tilde R a+c$ and
$L=Kx=\tilde K a$ for some real constant $c$ (arising
due the commutation relations among the canonical operators), some complex $n \times n$ {\em Hermitian}
matrix $\tilde R$ and some complex $m \times n$ ($m$ here again denotes the
number of input and output fields in and out of $G$) matrix
$\tilde K$. Here the term passive for such systems is physically
motivated. By the given definition, the Hamiltonian $H$ contains no terms of the
form $c_1a_j^2$, $c_2a_{k}^{*2}$, $c_3 a_{j}a_k$ and $c_4 a_j^* a_{k}^*$  and the coupling operator $L$ contains
no terms of the form $c_5 a_k^*$ (here $c_1,\ldots,c_5$ denote arbitrary complex constants), with the
indexes $j$ and $k$ running over $1,\ldots,n$. These terms are precisely the terms that require an external source
of quanta (i.e., an external pump beam) to realize (see, e.g., \cite{NJD08}) and cannot be implemented only by passive
components like mirrors, beamsplitters and phase shifters. The absence of such ``active'' terms is the reason we
refer to this class of systems as passive.

Let us write $\half a^{\dag}\tilde R a$ and $ \tilde K a$ in the following way:
\begin{eqnarray*}
\half a^{\dag}\tilde R a &=& \half \left[\begin{array}{cc}
a^{\dag} & a^T
\end{array}\right]\left[\begin{array}{cc} \half \tilde R & 0_{n \times n}\\ 0_{n \times n} & \half \tilde R^{\#}\end{array}\right]
\left[\begin{array}{c} a \\ a^{\#}\end{array}\right] -\frac{1}{4}\sum_{j=1}^{n}\tilde R_{jj}\\
&=&\half  x^T \Re\{\Sigma^{\dag}\tilde R\Sigma\}x-\frac{1}{4}\sum_{j=1}^{n}\tilde R_{jj}\\
\tilde K a &=& \tilde K \left[\begin{array}{cc} I & 0
\end{array}\right]\left[\begin{array}{c} a \\
a^{\#}\end{array}\right]=\tilde K\Sigma x.
\end{eqnarray*}
Therefore, $H=\half x^T R x$  with $R= \Re\{\Sigma^{\dag}\tilde
R\Sigma\}$, and $L=Kx$ with $K=\tilde K \Sigma$. It is easy to see
from this that for any passive system the block diagonal elements
$R_{jj}$ must be  {\em diagonal} matrices of the form $\lambda_j
I_{2}$ with $\lambda_j \in \Rbb$, the off diagonal block elements
$R_{jk}$ are real $2 \times 2$ matrices of the form
$R_{jk}=\left[\begin{array}{cc} \alpha_{jk} & \beta_{jk} \\
-\beta_{jk} & \alpha_{jk} \end{array} \right]$ for some real
numbers $\alpha_{jk}$ and $\beta_{jk}$, and the coupling matrix
$K_j$ to $x_j$ is of the form $K_j=[\begin{array}{cc} \gamma_j &
i\gamma_j \end{array}]$ for some complex number $\gamma_j$. One
would hope that a passive system can be synthesized using purely
passive components and the next theorem states that this is indeed
the case.

\begin{theorem}
\label{thm:passive-syn} Let $G_{\rm sys}=(S,L,H)$ be passive. Then the systems $\{G_j; j=1,\ldots,n\}$ constructed according to Corollary \ref{cor:qfn-synthesis} are all also passive.
\end{theorem}

The proof of the theorem can be found in Appendix \ref{sec:proof-passive-syn}.
We now conclude this section with an example of a passive system synthesis.

\begin{example}
Consider the realization problem of a passive two degrees of freedom linear
quantum stochastic system $G_{\rm sys}=(I,Kx,1/2 x^T R x)$
($x=(q_1,p_1,q_2,p_2)^T$) with
\begin{align*}
K &=[\begin{array}{cc} K_1 & K_2
\end{array}]=[\begin{array}{cccc} -3+i & -1-3i & 1 & i \end{array}]\;  (K_j \in \Cbb^{1 \times 2},\;j=1,2)
;\\
R &=\left[\begin{array}{cc} R_{11}& R_{12} \\ R_{12}^T & R_{22}
\end{array} \right]=\left[\begin{array}{cccc} 2 & 0 & 1 & 4 \\
0 & 2 & -4 & 1\\
1 & -4 & 1 & 0  \\
4 & 1 & 0 & 1 \end{array} \right]\; (R_{jk} \in \Rbb^{2 \times
2},\;j,k=1,2).
\end{align*}
Setting $\kappa_{12}=1$, $S_{12}=1$, $S_{21}=i$ and
$K_{12}=[\begin{array}{cc} \kappa_{12} &
i\kappa_{12}\end{array}]=[\begin{array}{cc} 1 & i \end{array}]$,
we obtain from Corollary \ref{cor:qfn-synthesis}, $K_{11}= K_1$,
$K_{22}=K_2$, $K_{21}=[\begin{array}{cc} 0.5-0.5i &  0.5+0.5i
\end{array}]$, $R_1 =0_{2 \times 2}$ and $R_2=0.5 I_{2}$. An
entirely passive optical circuit that realizes $G_{\rm sys}$ is
shown in Figure~\ref{fig:Gsys-passive-circuit}.

\begin{figure}%[h!]
\centering
\includegraphics[scale=0.5]{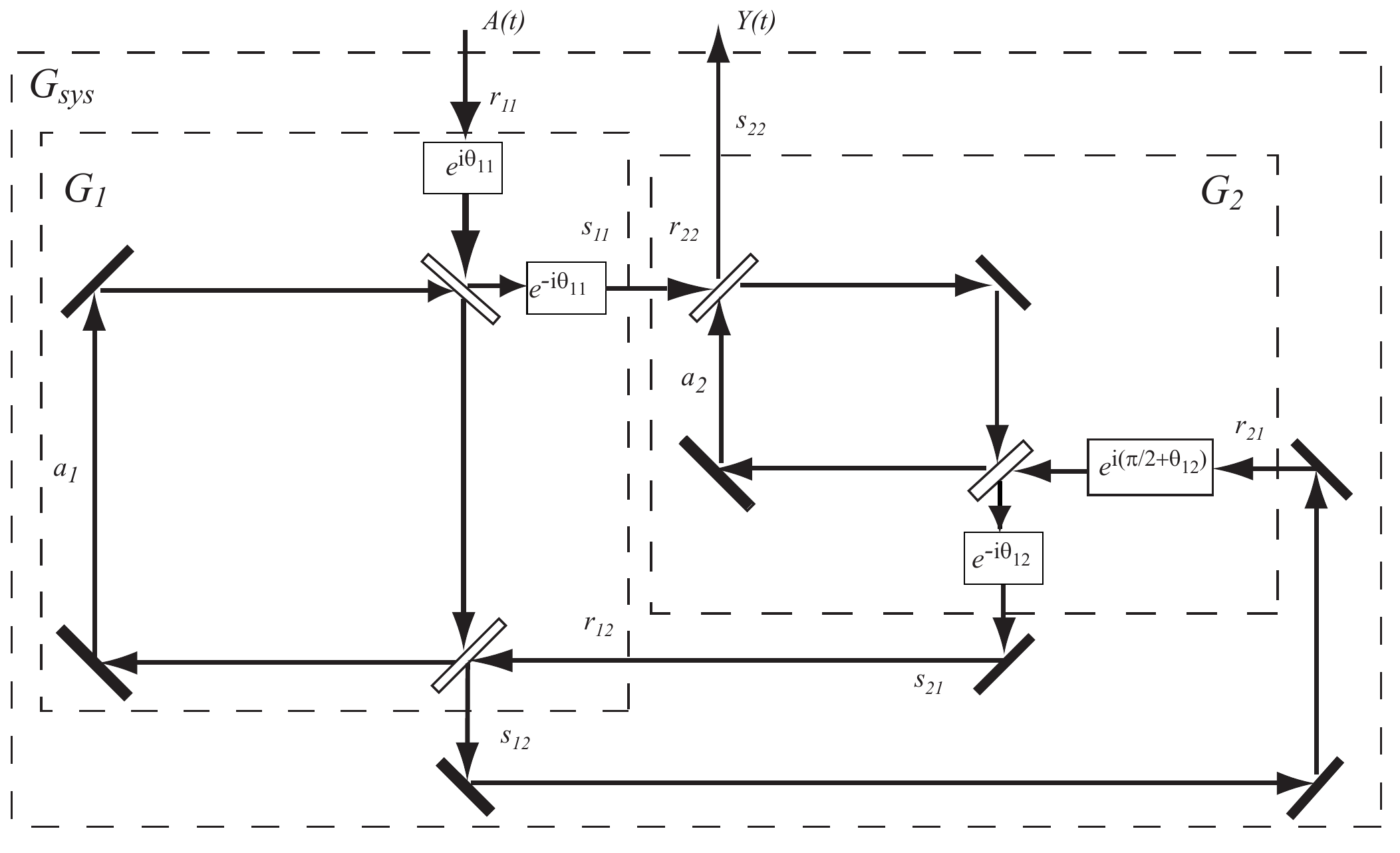}
\caption{A quantum optical circuit that realizes the passive system  $G_{\rm sys}$ of this example.
The circuit consists of only passive optical components: mirrors and phase shifters.
Here $\theta_{11}=\pi-\arctan(1/3)$ and $\theta_{21}=-\frac{\pi}{4}$, and the values of
all other parameters of devices in this circuit
can be determined according to \cite{NJD08}.}
\label{fig:Gsys-passive-circuit}
\end{figure}

\end{example}

\section{Conclusions}
\label{sec:conclu} In this paper we have developed a new method or algorithm
for systematically synthesizing an arbitrary linear quantum
stochastic system via a suitable quantum feedback network. In the
synthesis, all interactions between the oscillators constituting
the network are facilitated only by quantum fields. In particular,
it offers an alternative to the challenging task of implementing
direct bilinear interaction Hamiltonians that was required in the
approach of \cite{NJD08}. Moreover, it is shown that if the system is passive
then with the new algorithm it can be realized using only passive components.
It is clearly also possible to combine the present algorithm with that
of \cite{NJD08} to form a hybrid synthesis method.

Current interest in quantum information science, and quantum control in particular,
provides much impetus for extending network synthesis theory to the quantum domain as
a significant direction in which to further develop circuit and systems theory. The quantum synthesis results presented here could be of particular relevance for the design and systematic physical realization of photonics based monolithic linear open quantum circuits which may form an important sub-system of future continuous variable quantum information processing systems.

\appendix

\section{Proof of Theorem \ref{thm:qfn-construct}}
\label{sec:proof-qfn-construct} For this proof, it will be
convenient to interchange some rows and columns of the model
matrix $M$ to form another model matrix $\tilde M$ to avoid
complicated book keeping and thus reduce unnecessary clutter. 
Here by ``rows'' and ``columns'' we mean, respectively,
block rows and block columns of $M$ formed with respect to its specifed partitioning.
This interchange is as follows.

First, we permute rows of $M$ such that the first $2n-1$ rows from
top to bottom are the rows labelled (while column labels are kept
fixed as they are)
$s_{00},s_{12},s_{21},s_{13},s_{31},\ldots,s_{1n},s_{n1}$, the
next $2(n-2)$ rows  respectively are the rows labelled
$s_{23},s_{32},s_{24},s_{42},\ldots,s_{2n},s_{n2}$, the $2(n-3)$
rows after that are respectively the rows  labelled
$s_{34},s_{43},s_{35},s_{53},\ldots,s_{3n},s_{n3}$ and so on in
the same pattern until we get to the last $n$ rows that are
respectively those rows labelled $s_{11},s_{22},\ldots,s_{nn}$.
Call the intermediate matrix resulting from this row permutation
$\hat  M$. Then fixing the row labels of $\hat M$, we permute its
columns such that the first $2n-1$ columns from left to right are
respectively the columns of $\hat M$ labelled
$r_{00},r_{12},r_{21},r_{13},r_{31},\ldots,r_{1n},r_{n1}$, the
next $2(n-2)$ columns are respectively the columns labelled
$r_{23},r_{32},r_{24},r_{42},\ldots,r_{2n},r_{n2}$, the
 $2(n-3)$ rows after are respective the columns labelled
$r_{34},r_{43},r_{35},r_{53},\ldots,r_{3n},r_{n3}$, and so on in
the same pattern until the final $n$ columns that are respectively
the columns labelled $r_{11},r_{22},\ldots,r_{nn}$.
The resulting matrix after this permutation of columns is $\tilde M$.

It is important to note here that since the same permutation is
applied to the rows and columns, $M$ and $\tilde M$ are model
matrix representations of the same physical system, save for a
mere rearrangement of the ordering or indexing of the fields and
ports. That is to say that if $M$ is the model matrix of
$G=(S,L,H)$ then $\tilde M$ is the model matrix of $\tilde
G=(PSP^T,PL,H)$ for some suitable constant real permutation matrix $P$,
while it is clear that $G$ and $\tilde G$ are representations of
the same physical system. Thus with the same internal connections made,
a reduced model  matrix for $\tilde
M$ is also a reduced model matrix for $M$, up to a possible relabelling of uneliminated ports.

Let $\tilde L=PL$ and $\tilde S=PSP^T$. Then $\tilde L$ can be
partitioned as $\tilde L=(\tilde L_{\rm i}^T,\tilde L_{\rm
e}^T)^T$, where $\tilde L_{\rm i}$ is the first $n(n-1)+1$ rows of
$\tilde L$, while $\tilde L_{\rm e}$ is the last $n$ rows of
$\tilde L$. They are of the form:
\begin{eqnarray*}
\tilde L_{\rm i}&=&(L_{12}^T,L_{21}^T,L_{13}^T,L_{31}^T,\ldots,L_{1n}^T,L_{n1}^T,L_{23}^T,L_{32}^T,L_{24}^T,L_{42}^T,\ldots,L_{2n}^T,L_{n2}^T,\ldots,\\
&& L_{(n-2)(n-1)}^T,L_{(n-1)(n-2)}^T,L_{(n-2)n}^T,L_{n(n-2)}^T,L_{(n-1)n}^T,L_{n(n-1)}^T)^T\\
\tilde L_{\rm e}&=&(L_{11}^T,L_{22}^T,\ldots,L_{nn}^T)^T.
\end{eqnarray*}
Similarly, $\tilde S$ can be partitioned as
$\tilde S=\left[\begin{array}{cc} \tilde S_{\rm ii} & \tilde S_{\rm ie} \\
\tilde S_{\rm ei}& \tilde S_{\rm ee} \end{array} \right]$, with
$\tilde S_{\rm ii}$ and $\tilde S_{\rm ee}$ being block diagonal:
\begin{eqnarray*}
\tilde S_{\rm
ii}&=&\diag(S_{12},S_{21},S_{13},S_{31},\ldots,S_{1n},S_{n1},S_{23},S_{32},S_{24},S_{42},\ldots,S_{2n},S_{n2},\ldots,\\
&& S_{(n-2)(n-1)},S_{(n-1)(n-2)}, S_{(n-2)n},S_{n(n-2)},S_{(n-1)n},S_{n(n-1)})\\
\tilde S_{\rm ee}&=&\diag(S_{11},S_{22},\ldots,S_{nn}),
\end{eqnarray*}
and $\tilde S_{\rm ei}$ and $\tilde S_{\rm ie}$ both being zero
matrices. Then $\tilde M$ has the partitioning of the form (\ref{eq:in-out-partition}) by identifying $S_{\rm ii}$,
$S_{\rm ie}$, $S_{\rm ei}$, $S_{\rm ee}$, $L_{\rm i}$ and $L_{\rm e}$ with $\tilde S_{\rm ii}$, $\tilde S_{\rm ie}$, $\tilde S_{\rm ei}$, $\tilde S_{\rm ee}$ and $\tilde L_{\rm i}$ and $\tilde L_{\rm e}$, respectively.
The reduced model matrix resulting from the subsequent
simultaneous elimination of all internal edges $(s_{jk},r_{kj})$
$j,k=1,\ldots,n,j\neq k$, can be conveniently determined by using
the adjacency matrix $\eta$ defined by:
\begin{eqnarray*}
\eta &=& \diag\left(\left[\begin{array}{cc} 0 & I_{c_{12}} \\
I_{c_{21}}
& 0\end{array}\right], \left[\begin{array}{cc} 0 & I_{c_{13}} \\
I_{c_{31}} & 0\end{array}\right], \ldots, \right.\\
&&\quad \left. \left[\begin{array}{cc}
0 & I_{c_{1n}} \\ I_{c_{n1}} & 0\end{array}\right],\left[\begin{array}{cc} 0 & I_{c_{23}} \\ I_{c_{32}} & 0
\end{array}\right], \left[\begin{array}{cc} 0 & I_{c_{24}} \\ I_{c_{42}}
& 0\end{array}\right], \right.\\
&&\quad \left. \ldots, \left[\begin{array}{cc} 0 &
I_{c_{2n}} \\ I_{c_{n2}} & 0\end{array}\right],\ldots,  \left[\begin{array}{cc} 0 & I_{c_{(n-1)n}} \\
I_{c_{(n-1)n}} & 0\end{array}\right]\right).
\end{eqnarray*}
(Recall that $c_{jk}=c_{kj}$). Hence, according to Theorem \ref{thm:qfn-simul-elim}, the
reduced model matrix $\tilde M_{\rm red}$ obtained after
elimination of the internal edges $ \{(s_{jk},r_{kj});\,
j,k=1,\ldots,n,j\neq k\}$, has parameters given by
(recalling that
$\tilde S_{\rm ei}$ and $\tilde S_{\rm ie}$ are zero matrices):
\begin{eqnarray*}
\tilde S_{\rm red} &=& \tilde S_{\rm ee} + \tilde S_{\rm
ei}(\eta-\tilde S_{\rm ii})^{-1}\tilde S_{\rm
ie}= \tilde S_{\rm ee},\\
\tilde L_{\rm red}
&=& \tilde L_{\rm e}+\tilde S_{\rm ei}(\eta-\tilde S_{\rm ii})^{-1}\tilde L_{\rm i} = \tilde L_{\rm e}\\
\tilde H_{\rm red}&=& \sum_{k=1}^n H_k + \sum_{j={\rm i,e}}
\Im\{\tilde L_j^{\dag}\tilde S_{j{\rm i}}(\eta-\tilde S_{\rm
ii})^{-1}\tilde L_{\rm i}\} \\
&=& \sum_{k=1}^n H_k +  \Im\{\tilde L_{\rm i}^{\dag}\tilde S_{{\rm
ii}}(\eta-\tilde S_{\rm ii})^{-1}\tilde L_{\rm i}\} \\
&=& \sum_{k=1}^n H_k +  \Im\{\tilde L_{\rm i}^{\dag}\eta (\eta-\tilde
S_{\rm ii})^{-1}\tilde L_{\rm i}\} \\
&=& \sum_{k=1}^n H_k + \sum_{j=1}^{n-1} \sum_{k=j+1}^{n}
\Im\biggl\{[\begin{array}{cc} L_{jk}^{\dag} &
L_{kj}^{\dag}\end{array}]  \left[\begin{array}{cc} I & -S_{jk} \\
-S_{kj} & I \end{array} \right]^{-1}\left[\begin{array}{c}
L_{jk} \\ L_{kj} \end{array} \right]\biggr\}
\end{eqnarray*}
Since $M$ and $\tilde M$ are model matrix representations of the
same physical system and the external fields have the same
ordering and labelling in both representations, the reduced model
matrix of $M_{\rm red}$ and $\tilde M_{\rm red}$ of $M$ and
$\tilde M$, respectively, after elimination of internal edges
$(s_{jk},r_{kj})$, coincide. Hence, also the linear quantum
stochastic systems $G_{\rm red}$ and $\tilde G_{\rm red}$
associated with $M$ and $\tilde M$, respectively, coincide. This
completes the proof. \hfill $\Box$

\section{Proof of Lemma \ref{lem:direct-coupling}}
\label{sec:proof-direct-coupling}
We begin by noting that
\begin{eqnarray*}
\Im\{\frac{S_{12}}{1-S_{12}S_{21}}K_1^{\dag}K_2+ \frac{S_{21}}{1-S_{12}S_{21}}K_1^T
 K_2^{\#}\} =\frac{1}{2i}[\begin{array}{cc} -K_1^{\dag}\Delta^* & K_1^T\Delta \end{array}]\left[
\begin{array}{c}  K_2 \\ K_2^{\#}\end{array} \right],
\end{eqnarray*}
with $\Delta=\frac{S_{21}}{1-S_{21}S_{12}}-\frac{S_{12}^*}{1-S_{21}^*S_{12}^*}=2\frac{S_{21}-S_{12}^*}{|1-S_{21}S_{12}|^2}$
(exploiting the fact that $S_{12}S_{12}^*=1=S_{21}S_{21}^*$). Now set $K_1=[\begin{array}{cc}
\kappa & i\kappa
\end{array}]$ for an arbitrary non-zero real constant $\kappa$, and note that $S_{21}S_{12} \neq 1$ implies
that $\Delta \neq 0$ and:
\begin{eqnarray*}
[\begin{array}{cc} -K_1^{\dag}\Delta^* & K_1^T\Delta
\end{array}]^{-1} 
&=&\left[\begin{array}{cc}
-\kappa \Delta^* & \kappa\Delta \\
i\kappa \Delta^* & i\kappa \Delta
\end{array}\right]^{-1} = -\frac{1}{2i\kappa^2 |\Delta|^2}\left[\begin{array}{cc}
i\kappa \Delta & -\kappa\Delta \\
-i\kappa \Delta^* & -\kappa \Delta^*
\end{array}\right],
\end{eqnarray*}
and therefore for any real matrix $V$, $2i[\begin{array}{cc}
-K_1^{\dag}\Delta^* & K_1^T\Delta
\end{array}]^{-1}V=\left[\begin{array}{c} Z \\ Z^{\#}\end{array}\right] $ for some
complex row vector $Z$. Therefore, given any $R$ we see that we
may solve the equation
$$
[\begin{array}{cc} -K_1^{\dag}\Delta^* & K_1^T\Delta
\end{array}]\left[
\begin{array}{c}  K_2 \\ K_2^{\#}\end{array} \right]=2iR,
$$
for $K_2$ and this solution is as given in the statement of the
corollary.

Alternatively, we could also have started by setting
$K_2=[\begin{array}{cc} \kappa & i\kappa \end{array}]$ and
analogously solving for $K_1$ for a given $R$. It is then an easy
exercise that the solution for $K_1$ in this case is as stated in
the corollary. \hfill $\Box$

\section{Proof of Corollary \ref{cor:qfn-synthesis}}
\label{sec:proof-qfn-synthesis} With $c_{jk}$, $S_{jk}$, $R_{jk}$
and $K_{jk}$, $j,k=1,\ldots,n$, as defined in the statement of the corollary, from
Theorem \ref{thm:qfn-construct} and Lemma
\ref{lem:direct-coupling} we have that $S_{\rm red}=I_{nm}$,
$L_{\rm red}=(L_{11}^T,L_{22}^T,\ldots,L_{nn}^T)^T$ with
$L_{jj}=K_jx_j$, and
\begin{eqnarray*}
H_{\rm red}&=& \sum_{j=1}^n H_j+ \sum_{j=1}^{n-1}\sum_{k=j+1}^{n} \Im\{
[\begin{array}{cc} L_{jk}^{\dag} & L_{kj}^{\dag} \end{array}] \left[\begin{array}{cc} 1 & -S_{jk} \\ -S_{kj} & 1 \end{array} \right]^{-1}
\left[\begin{array}{c} L_{jk} \\ L_{kj} \end{array}\right]\}.
\end{eqnarray*}
Expanding, we have:
\begin{eqnarray*}
H_{\rm red} &=& \half \sum_{j=1}^n x_j^T R_{j} x_j + \sum_{j=1}^{n-1}\sum_{k=j+1}^{n} \Im\biggl\{
\frac{1}{1-S_{jk}S_{kj}} \cdot \\
&&\;(L_{jk}^{\dag}L_{jk}+S_{jk}L_{jk}^{\dag}L_{kj}+ S_{kj}L_{kj}^{\dag}L_{jk}+
L_{kj}^{\dag}L_{kj}) \biggr\}\\
%&=&\half \sum_{j=1}^{n} x_j^T \biggl(R_{j}+2\sum_{k=1,k \neq j}^{n} \Im\{\frac{K_{jk}^{\dag}K_{jk}}{1-S_{jk}S_{kj}}\}\biggr) x_j + \\
%&&\quad x_j^T\sum_{j=1}^{n-1}\sum_{k=j+1}^n
%\Im\{\frac{S_{jk}}{1-S_{jk}S_{kj}} K_{jk}^{\dag}K_{kj}+ \\
%&&\; \frac{S_{kj}}{1-S_{jk}S_{kj}} K_{jk}^T K_{kj}^{\#} \}x_k\\
&=& \half \sum_{j=1}^{n} x_j^T \biggl(R_{j}+ 2 {\rm sym}\biggl(
\sum_{k=1,k \neq j}^{n} \Im\{\frac{K_{jk}^{\dag}K_{jk}}{1-S_{jk}S_{kj}}\}\biggr)\biggr) x_j\\
&&\; + \sum_{j=1}^{n-1}\sum_{k=j+1}^n
x_j^T\Im\{\frac{S_{jk}}{1-S_{jk}S_{kj}} K_{jk}^{\dag}K_{kj}+   \frac{S_{kj}}{1-S_{jk}S_{kj}} K_{jk}^T K_{kj}^{\#} \}x_k\\
&=& \half \sum_{j=1}^{n} x_j^T  R_{jj} x_j + \sum_{j=1}^{n-1}\sum_{k=j+1}^n
x_j^T\bigl(R_{jk}-  \Im\{K_{j}^T K_{k}^{\#}\}\bigr)x_k\\
&=& \half x^T R x - \sum_{j=1}^{n-1}\sum_{k=j+1}^{n}x_j^T
\Im\{K_{j}^T K_{k}^{\#}\}x_k,
\end{eqnarray*}
where ${\rm sym}(A)=\half(A+A^T)$, and $R=[R_{jk}]_{j,k=1,\ldots,n}$ and $R_{kj}=R_{jk}^T$. From
this it is clear that using the concatenation product we can
decompose $G_{\rm red}$ as $G_{\rm red}=(0,0,H_{\rm red}) \boxplus
\boxplus_{j=1}^{n} (I_m,L_{jj},0)$. Let $G_{{\rm red},0}=(0,0,H_{\rm red})$
and $G_{{\rm red},j}=(I_m,L_{jj},0)$, $j=1,\ldots,n$. Now, using the series product
rule (cf. Section \ref{sec:linear-summary}), we easily compute that
\begin{eqnarray*}
G_{\rm net}&=& G_{{\rm red},0} \boxplus (G_{{\rm red},n} \triangleleft \ldots \triangleleft  G_{{\rm red},2} \triangleleft G_{{\rm red},1}) \\
&=& \biggl(0,0,\half x^T R x -\sum_{j=1}^{n-1}\sum_{k=j+1}^{n}x_j^T \Im\{K_j^T K_k^{\#}\} x_k\biggr)  \boxplus \\
&& \biggl(I_{m},[\begin{array}{cccc} K_1 & K_2 & \ldots & K_n
\end{array}]x,
\sum_{j=1}^{n-1}\sum_{k=j+1}^{n}x_j^T   \Im\{K_j^T K_k^{\#}\} x_k\biggr) \\
&=& \biggl( I_{m},[\begin{array}{cccc} K_1 & K_2 & \ldots & K_n
\end{array}]x, \half x^T R x\biggr).
\end{eqnarray*}

Therefore, $G_{\rm net}$ realizes a linear quantum stochastic system  with parameters $S_{\rm net}$, $L_{\rm net}$ and
$H_{\rm net}$, as claimed.\hfill $\Box$

\section{Proof of Theorem \ref{thm:passive-syn}}
\label{sec:proof-passive-syn} By the passivity of $G_{\rm sys}$, we immediately see that
 $K_{jj}=K_j$ is already of the form required for passivity of $G_{j}$, and the matrix
$R_{jk}- \Im\{K_j^T K_k^{\#}\}$ is a $2 \times 2$ real matrix of
the form $\left[ \begin{array}{cc} \alpha'_{jk} & \beta'_{jk} \\
-\beta'_{jk} & \alpha'_{jk} \end{array} \right]$ for some
$\alpha'_{jk},\beta'_{jk} \in \Rbb$, whenever $j \neq k$. Let
$\kappa_{jk}$, $S_{jk}$ and $S_{kj}$ be chosen according to
Corollary  \ref{cor:qfn-synthesis}. Set $K_{jk}$ according to the
first equality of (\ref{eq:K-coupling-1}). Then some
straightforward algebra shows that $K_{kj}$ given by the second
equality of (\ref{eq:K-coupling-1}) is of the form
$K_{kj}=[\begin{array}{cc} \gamma'_{kj} & i\gamma'_{kj}
\end{array}]$ for some $\gamma'_{kj} \in \Cbb$, just like
$K_{jk}$. The same holds true if one chooses the alternative of
setting $K_{kj}$ according to the first equality of
(\ref{eq:K-coupling-2}) and computing $K_{jk}$ according to the
second equality of (\ref{eq:K-coupling-2}). Finally, given this
special form of $K_{jk}$, it is easily inspected that ${\rm
sym}\bigl(\Im\{\frac{1}{1-S_{jk}S_{kj}}
K_{jk}^{\dag}K_{jk}\}\bigr)$ is a diagonal matrix of the form
$\lambda_{jk} I_{2}$ for some $\lambda_{jk} \in \Rbb$ for all
$j,k,j\neq k$ (recall that ${\rm sym}(A)=\half (A+A^T)$), and
since $R_{jj}$ is also diagonal of this form (again from the
passivity of $G_{\rm sys}$) we have that
$R_j=R_{jj}-2\sum_{k=1,k\neq j}^n {\rm
sym}\bigl(\Im\{\frac{1}{1-S_{jk}S_{kj}}
K_{jk}^{\dag}K_{jk}\}\bigr)$ is again of the same form, for all
$j$. Therefore,  it now follows that each sub-system $G_j$ of
Section \ref{sec:main-results} with parameters determined
according to Corollary \ref{cor:qfn-synthesis} is passive. This
completes the proof.

\bibliographystyle{ieeetran}
\bibliography{ieeeabrv,rip,mjbib2004,irpnew}

\end{document}